\documentclass[10pt,twocolumn]{article}
\usepackage{latex8}

\usepackage{latexsym}
\usepackage{amsmath}
\usepackage{amssymb}
\usepackage{makeidx}

\newcommand{\CC}{C\nolinebreak\hspace{-.05em}\raisebox{.4ex}{\tiny\bf +}\nolinebreak\hspace{-.10em}\raisebox{.4ex}{\tiny\bf +}}
\newcommand{\Cc}{C\nolinebreak\hspace{-.05em}$+$\nolinebreak\hspace{-.05em}$-$}

\usepackage{float}
\usepackage{alltt}

\ifx\pdfoutput\undefined
\usepackage[hypertex]{hyperref}
\else
\usepackage[pdftex,colorlinks=true,hyperindex,backref,hypertexnames=false]{hyperref}
\usepackage[pdftex]{graphicx}
\fi

\hypersetup{%
  pdftitle={Trading off Complexity for Expressiveness in Programming Languages: Visions and Preliminary Experiences}
  pdfsubject={Software engineering},
  pdfkeywords={Programming languages, source-to-source program transformation, compilers, translators, embedded systems},
  bookmarksopen=true,
  bookmarksnumbered=true,
  pdfstartview={FitH},
  urlcolor=blue,
}%

\def\UNDSCR{\_}

\begin{document}

\title{Trading off Complexity for Expressiveness in Programming Languages:
       Visions and Preliminary Experiences}

\author{Vincenzo De Florio and Chris Blondia \vspace*{3pt} \\
        University of Antwerp\\
        Department of Mathematics and Computer Science\\
        Performance Analysis of Telecommunication Systems group\\
        Middelheimlaan 1, 2020 Antwerp, Belgium \vspace*{3pt} \\
\emph{and}\\
        Interdisciplinary Institute for Broadband Technology (IBBT)\\
        Gaston Crommenlaan 8, 9050 Ghent-Ledeberg, Belgium}

\maketitle

\begin{abstract}
When programming resource-scarce embedded smart devices, the designer
often requires both the low-level system programming features of a language such as C and 
higher level capability typical of a language like Java.
The choice of a particular language typically implies trade offs between conflicting
design goals such as performance, costs, and overheads. The large variety
of languages, virtual machines, and translators provides the designer with a dense trade off space,
ranging from minimalistic to rich full-fledged approaches, but once a choice is
made it is often difficult for the designer to revise it. In this work
we propose
a system of light-weighted and modular 
extensions as a method to flexibly
reshape the target programming language as needed, adding only those 
application layer features that match the current design goals.
In so doing complexity is made transparent, but not hidden:
While the programmer can benefit of higher level constructs, the
designer can deal with modular building blocks each characterized by
a certain algorithmic complexity and therefore each accountable for a given
share of the overhead.
As a result the designer is given a finer control on the amount of resources
that are consumed by the run-time executive of the chosen programming
language.
\end{abstract}

\Section{Introduction}
The July 2010 Tiobe 
Programming Community index~\cite{Tiobe}, ranking programming languages 
according to their matching rate in several search engines, sets C as 
the second most popular programming language, barely 0.2\% less than 
Java. 
C's object-oriented counterpart \CC{} is third but quite further 
away (18.48\% vs. 10.469\%). Quite remarkably, C was still 
``programming language of the year'' for Tiobe in 2008, exhibiting that 
is the highest rise in ratings in that year, as Java was in 2005.
Both quite successful and wide-spread, C and Java represent
two extremes of a 
spectrum of programming paradigms ranging from system-level to service-level development.
Interestingly enough,
in C complexity is mostly in the application layer, as its run-time executive is typically
very small~\cite{KeRi2}; in Java, on the other hand, non-negligible
complexity comes also with an often rich execution environment (EE). 
The latter comprises a virtual machine and advanced features such as autonomic
garbage collection.
The only way to trade off the EE complexity for
specific services is then by switching to another EE.

Various EE's are
available, developed by third parties to match specific classes of target platforms.
Fine-tuning the EE is also possible, e.g. in Eclipse; and of course it is also
possible to go for a custom implementation. In general though
the amount and the nature of the EE complexity is
hidden to the programmer and the designer---after all, it is 
the very same nature of Java as a portable
programming language that forbids to exploit such knowledge.

Though transparent, such hidden complexity is known to have an impact on several aspects, including
overhead, real-timeliness, deterministic behavior, and security~\cite{mw09.12}.
In particular, when a computer system's expected activity
is well defined and part of that system's
quality of service---as it is the case e.g. for 
real-time embedded systems---then any task with unknown algorithmic complexity or exhibiting
non-deterministic behavior might simply be unacceptable. As an example,
a run-time component asynchronously recollecting unused memory,
though very useful in itself, 
often results in asynchronous, unpredicted system activity affecting
e.g. the processors and the memory system---including caches.
Taking asynchronous tasks such as this into account would 
impact negatively on the analysis of worst-case execution times
and consequently on costs as well.

In what follows we propose an alternative---in a sense, an \emph{opposite}---direction:
Instead of stripping functionality from Java to best match a given target platform,
we chose to add functionality to C to compensate for lack of expressiveness
and linguistic support. More specifically, in our approach, C with its
minimalistic run-time executive becomes a foundation on top of which the
designer is made able to easily lay a system of modular linguistic extensions. 
By doing so
the above mentioned partitioning of complexity is not statically defined and unchangeable,
but rather it becomes
revisable under the control of the designer. Depending on the
desired linguistic features and the overhead permitted by the target platform
as well as by mission and cost constraints, our approach allows the
programming language to be flexibly reshaped. 
This is because our approach employs
well-defined ``complexity containers'', each of which provides limited specific
functions and each of which is characterized by well-defined complexity and overhead.
Syntactic features and EE functions
are weaved together under the control of the designer, resulting in bound and known
complexity. A dynamic trade off between 
complexity and expressiveness can then be achieved and possibly revised in later development
stages
or when the code is reused on a different platform. In principle such combination
of transparent functionality and translucent complexity should also reduce the
hazards of unwary reuse of software modules~\cite{Lev95}.

The current version of our system is simply a proof-of-concepts; in 
particular the control of the augmentations is still manual, 
which makes our prototypical implementation far from being perfect.
Assisted, automatic, or even intelligent 
assembling of our extensions is the matter of our current research.

The structure of this paper is as follows: In Sect.~\ref{s:bascom} we 
introduce a number of ``basic components'' respectively implementing extensions
for context awareness, for autonomic 
data integrity, and for event management. In 
Sect.~\ref{s:combin} we discuss how we built such components and how 
they may be dynamically recombined and thus give raise to specific 
language variants. Section~\ref{s:evalue} introduces a case study and 
some preliminary evaluation. Our conclusions are finally produced in 
Sect.~\ref{s:conclu}.

\Section{Basic Components}\label{s:bascom}

This section introduces three basic components of our approach: 
Linguistic support to context awareness 
(Sect.~\ref{ss:reflec}), adaptive redundancy management 
(Sect.~\ref{ss:redund}), and 
application-level management of cyclic events (Sect.~\ref{ss:schedu}).
In all three cases the syntactical extension instruments the memory access
operations on certain variables.

\SubSection{Context Awareness Component}\label{ss:reflec}

Context awareness (CA)
is defined here as the ability to expose certain properties and
accordingly react to certain conditions in a 
transparent and intuitive way. Such properties may be endogenous or 
exogenous, and are called herein ``the context''. Endogenous context 
describes properties of computer-specific processes: Structure and state 
of the software components, hardware properties, operating system state, 
policies supported by the EE---to name but a few 
examples. Exogenous context is properties regarding the processes 
occurring between a computer system and the physical world. More 
specifically, it is ``any information that characterizes a situation 
related to the interaction [of a computer system with] with humans, 
applications and the surrounding environment''~\cite{Dey01}.

In the rest of this section we focus
on linguistic support to CA. This may take different forms
depending on programming language design choices. 
Linguistic support to CA influences adaptability, which we
define here as the ability to structure one's function 
in accordance with a subset of the current endogenous and exogenous 
context conditions. Such subset represents a choice of context variables 
that are deemed as ``sensible enough'' to steer optimally the function 
of the system. Linguistic support to CA is the fundamental building block---at 
application layer---to build open and ``self-$\star$'' systems, i.e. 
flexible, adaptive systems able to autonomically re-optimize themselves 
in the face of changes. Many high level programming languages 
support CA via e.g. computational reflection, composition filters, or
aspects. Lacking 
such linguistic support it becomes more difficult and error-prone to 
design e.g. ambient-intelligent embedded devices. C has no built-in 
support for context awareness, which means that the designers requiring 
such service need to rely on ``external'' support, e.g. 
via middleware~\cite{mw14}.

What we call the CA component of our architecture 
is a translator that filters an ``augmented C'' source code producing a 
standard C source code. Such output code makes 
use of a thread and the methods in an external library. The translator 
intercepts occurrences of specific variables that are interpreted as 
access points to actuators and sensors, in a way similar to the one 
described in~\cite{DB07d}. Actuators are managed as overloaded 
assignment operators in \CC{}: Writing to an actuator variable
triggers a
side-effect, defined as a user-selected method call. Sensors are managed 
through threads, which transparently update shared memory locations with 
the current value of a context property, e.g. the current state of a 
watchdog timer thread, reified as a value fitting in a C variable of 
some type. Available sensors reflect context information such as the 
amount of CPU currently being used or the state of external components,
e.g. media players or watchdog timers. Guarded functions asynchronously
evaluate expressions on the sensors and are executed when their guards
become true. 

The most notable difference between this approach and e.g. 
the one in~\cite{DB07d}
is that sensors and actuators can be represented here as 
dynamically growing arrays that are addressable by domain-specific 
indices. As an example, \textsf{linkbeacons} is an array of ``objects'' 
(actually, structures) that represent Medium Access Control layer 
properties of mobile ad-hoc network peers. A new object comes to
life dynamically each time a new peer comes in proximity. When a peer 
node falls out of range, the corresponding object becomes ``stale''
until its node becomes reachable again. The \textsf{linkbeacons} array is addressed by 
strings representing the MAC address of peer nodes. Array 
\textsf{linkbeacons} reflects a number of properties, including the 
number of MAC beacons received by a peer node during the last 
``observation period'' (defined in our experiments as sixty seconds) or 
the number of periods elapsed without receiving at least one beacon from 
a certain node.

Similarly, array \textsf{linkrates} returns Network layer properties of 
peers in proximity---in particular, it returns the estimated bandwidth 
between the current node and the addressed one. 

\begin{figure}[t]
\centerline{\includegraphics[width=0.5\textwidth]{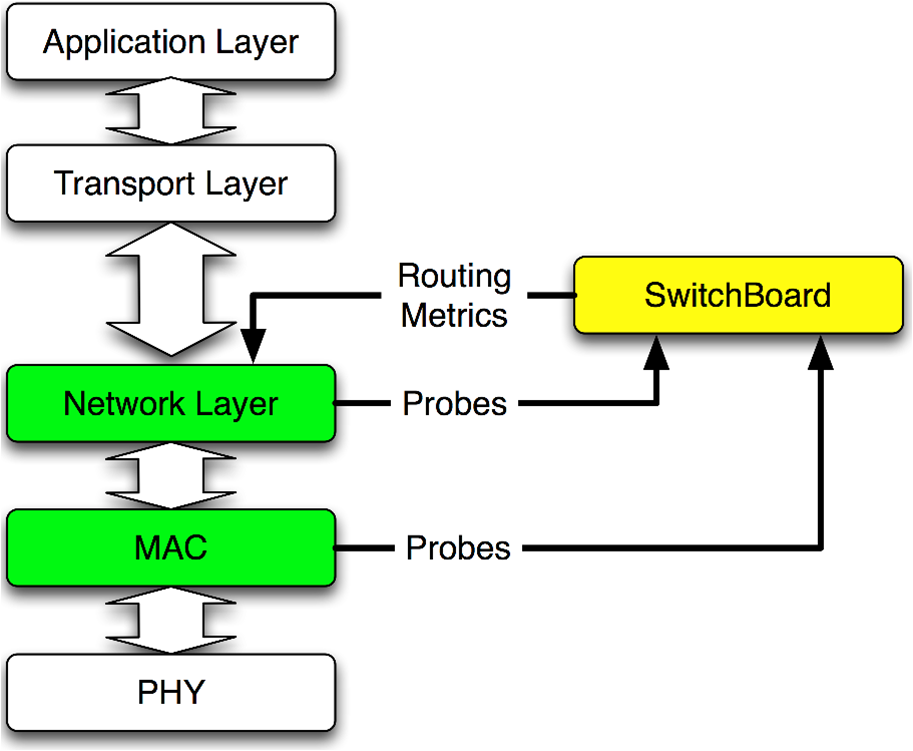}}
\caption{Computational reflection is used to expose MAC and Network layer properties via
  the \textsf{linkbeacons} and \textsf{linkrates} probes. The produced information is used 
  to compute routing metrics.}
\label{f:switch}
\end{figure}

The above mentioned arrays are currently being 
used in our research group to set up sort of cross-layer 
``switchboards'' able to perform optimizations such as MAC-aware IP 
routing in mobile ad-hoc networks (see Fig.~\ref{f:switch}).

\begin{figure}[t]
\centerline{\includegraphics[width=0.55\textwidth]{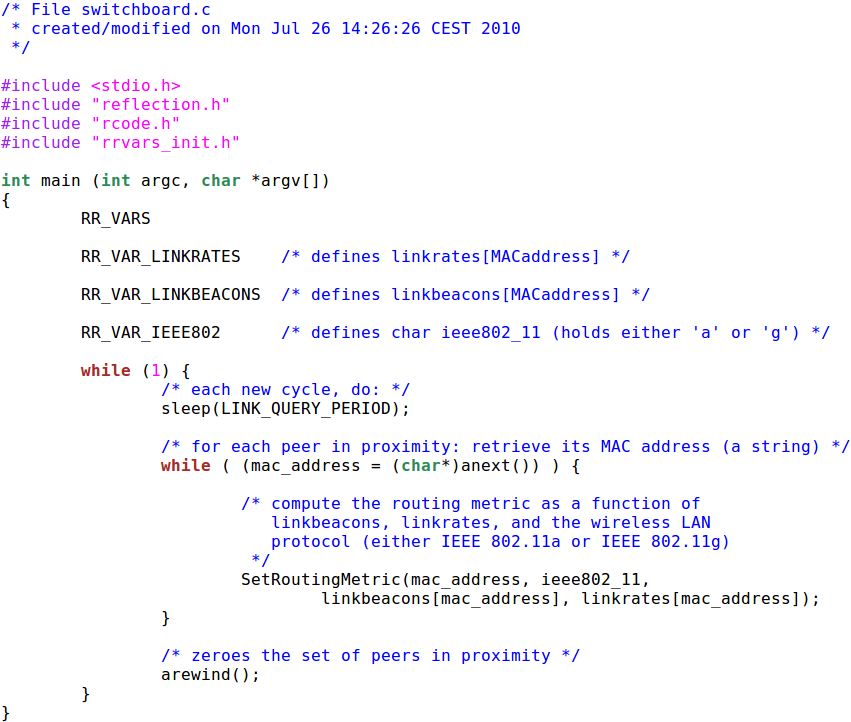}}
	\caption{The ``augmented C'' code of our cross-layer switchboard.}
	        \label{f:switchcode}
\end{figure}

As can be seen in Fig.~\ref{f:switchcode}, the program used to steer 
this cross layer optimization is quite simple: Every new observation 
cycle, the program retrieves the MAC addresses of the peers in proximity 
via a simple function call (\textsf{anext}) and then requests
to adjust the routing metric using the above mentioned 
arrays. The actual adjustments to the routing protocol are 
carried out through a Click~\cite{click:tocs00} script.

\SubSection{Adaptive Redundancy Component}\label{ss:redund}


Another important service that is typically missing in conventional 
programming languages such as C is transparent data replication. As 
embedded systems are typically streamlined platforms in which resources 
are kept to a minimum in order to contain e.g. costs and power consumption, 
hardware support to memory error detection is often missing. When such 
embedded systems are mission critical and subjected to unbound levels of 
electromagnetic interference (EMI), it is not uncommon to suffer from 
transient failures. As an example, several Toyota models recently experienced 
unintended acceleration and brake problems. Despite Toyota's official 
communications stating otherwise, many researchers and consultants are 
suggesting this to be just another case of EMI-triggered 
failures~\cite{toyo1,toyo2,toyo3}. More definitive evidence exists that EMI 
produced by personal electronic devices does affect electronic controls 
in modern aircrafts~\cite{PeGe96}, as it is the case for control 
apparatuses operating in proximity of electrical energy 
stations as well~\cite{DeBo98}. 

\begin{figure*}[t]
\centerline{\includegraphics[width=0.85\textwidth]{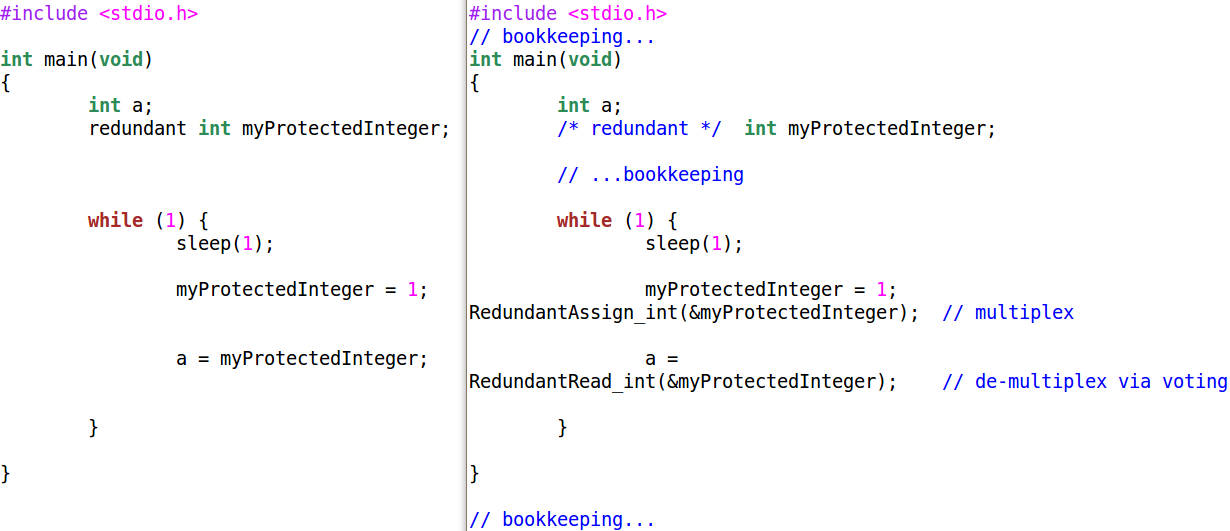}}
\caption{A simple example of use of redundant variables.
An ``extended C'' source code that accesses a redundant variable
(left-hand image) and an excerpt from the translation in plain C 
(right-hand picture) are displayed.}\label{f:ex1}
\end{figure*}

Whenever EMI causes unchecked memory corruption, a common strategy is
to use redundant data structures~\cite{TaMB80}: Mission-critical data structures
are then ``protected'' by replication and voting and 
through redoing~\cite{DeBo98}.
Our adaptive redundancy component is just a filter that allows the
user to tag certain variables as being ``redundant''. The filter transparently
replicates those variables according to some policy (for instance,
in separate ``banks'') and then catches memory accesses to those variables.
Write accesses are multiplexed and store their ``rvalues''~\cite{KeRi2}
in each replica, while read accesses are demultiplexed via a majority
voting scheme. Figure~\ref{f:ex1} summarizes this via a simple example.

In some cases, for instance when the application is
cyclic and constantly re-executed as in~\cite{DSN03}, the behavior
of the voting scheme can be monitored and provide an estimation
of the probability of failure: As an example, if
the errors induced by EMI are affecting a larger and larger amount
of replicas, then this can be interpreted as a symptom
for the imminent failure of the voting scheme due to the
impossibility to achieve a majority. Detecting this and assessing the
corresponding risk of voting failure allows the amount of replicas to be
transparently and autonomically adjusted, e.g. as described in~\cite{DF10a}.

\SubSection{Cyclic Methods Component}\label{ss:schedu}
As observed in~\cite{DeFBl10b}, linguistic constructs
such as \textsf{repeat periodically},
\textsf{at time $t$ send heartbeat},
\textsf{at time $t$ check whether message $m$ has arrived},
or \textsf{upon receive},
are often used to produce
pseudo-code for distributed protocols.
The lack of those constructs in a language such as C led us in the past
to implement a library of so-called ``time-out objects''.
Such objects postpone an associated function call by a
user-defined amount of time. In~\cite{DeFBl10b} we showed how this permits to
implement the above constructs by
converting time-based events into
message arrivals or signal invocations.
In the cited paper we also proposed some preliminary
``syntactic sugar'' to ease up the use of such objects in C.
Table~\ref{t:examp} is a simple example of how our time-out objects could be
used to define and control two ``cyclic methods,'' i.e., functions that
are executed by the run-time system every new user-defined cycle.

In the experience reported in this paper we capitalized on our previous
achievements and designed an extension that facilitates the definition
of cyclic methods.

\begin{table}[t]
\begin{sf}
\begin{small}
\begin{tabbing}
{\bf 001} \= tom\UNDSCR{}declare(\&t1, \= 000000000 \= 0000 \kill\\
{\bf 1.}\>  /* declarations */\\
        \> TOM *tom; timeout\UNDSCR{}t t1, t2;\\
        \> int PeriodicMethod1(TOM*), PeriodicMethod2(TOM*);\\
        \\
{\bf 2.}\> /* definitions */\\
        \> tom $\leftarrow$ tom\UNDSCR{}init();\\
        \> tom\UNDSCR{}declare(\&t1, TOM\UNDSCR{}CYCLIC, TOM\UNDSCR{}SET\UNDSCR{}ENABLE, \\
        \>                 \> TIMEOUT1, SUBID1, DEADLINE1);  \\
        \> tom\UNDSCR{}set\UNDSCR{}action(\&t1, PeriodicMethod1); \\
        \> tom\UNDSCR{}declare(\&t2, TOM\UNDSCR{}CYCLIC, TOM\UNDSCR{}SET\UNDSCR{}ENABLE, \\
        \>                 \> TIMEOUT2, SUBID2, DEADLINE2);  \\
        \> tom\UNDSCR{}set\UNDSCR{}action(\&t2, PeriodicMethod2); \\
        \\
{\bf 3.}\> /* insertion */\\
        \> tom\UNDSCR{}insert(tom, \&t1), tom\UNDSCR{}insert(tom, \&t2); \\
        \\
{\bf 4.}\> /* control */\\
        \> tom\UNDSCR{}disable(tom, \&t2); \\
        \> tom\UNDSCR{}set\UNDSCR{}deadline(\&t2, NEW\UNDSCR{}DEADLINE2); \\
        \> tom\UNDSCR{}renew(tom, \&t2); \\
        \> tom\UNDSCR{}delete(tom, \&t1);
\end{tabbing}
\end{small}
\end{sf}
\caption{Example of usage of the TOM time-out management class. In {\bf 1.} a time-out
list pointer and two time-out objects are declared, together with two
alarm functions. In {\bf 2.} the time-out list and the time-outs
are initialized.
Insertion is carried out in
{\bf 3.} In {\bf 4.}, 
time-out {\sf t2} is disabled; its deadline
is changed; {\sf t2} is restarted;
and finally, time-out {\sf t1} is deleted.}
\label{t:examp}
\end{table}

Table~\ref{t:examp2} shows the syntax of our extension.
In short, the extension allows the user to specify a dummy
member, \textsf{Cycle}, for those methods that have been tagged with
attribute
\textsf{cyclic\_t}. Every \textsf{Cycle} milliseconds our extension
executes a new instance of the corresponding method---irrespective
of the fact that previous instances are still running or otherwise.

\begin{table}[t]
\begin{sf}
\begin{small}
\begin{tabbing}
{\bf 001} \= tom\UNDSCR{}declare(\&t1, \= 000000000 \= 0000 \kill\\
{\bf 1.}\>  /* declarations */\\
	\> \textbf{cyclic\_t} int PeriodicMethod1(TOM*);\\
	\> \textbf{cyclic\_t} int PeriodicMethod2(TOM*);\\
        \\
{\bf 2.}\> /* definitions: unnecessary */\\
        \\
{\bf 3.}\> /* insertion */\\
	\> PeriodicMethod1\textbf{.Cycle} =  DEADLINE1; \\
	\> PeriodicMethod2\textbf{.Cycle} =  DEADLINE2; \\
        \\
{\bf 4.}\> /* control */\\
	\> PeriodicMethod2\textbf{.Cycle} = NEW\UNDSCR{DEADLINE2}; \\
	\> PeriodicMethod1\textbf{.Cycle} =  0;
\end{tabbing}
\end{small}
\end{sf}
\caption{The new syntax for the example of Table~\ref{t:examp}. Two simple constructs
are introduced---bold typeface is used to highlight their occurrences in this example.}
\label{t:examp2}
\end{table}

\Section{Putting Things Together}\label{s:combin}
In previous section we introduced a number of components each of which
can be used to augment plain C with extra features. In the rest of
this section we briefly describe the general design principles behind these
components (Sect.~\ref{ss:genpri}) and then we introduce our current simplistic
approach to combine them together (Sect.~\ref{ss:combin}).

\SubSection{General Design Principles}\label{ss:genpri}
The key principle of our approach is that
of source-to-source program transformation through a set of
independent and interchangeable extensions, each adopting a set of
orthogonal (that is, non-overlapping) syntaxes\footnote{In some
	cases there might arise conflicts when the same C entity is referred
	in two or more extensions. A practical example of this is
	a same variable being addressed by two extension, as it is
	the case for \textsf{watchdog} in Sect.~\ref{s:evalue}.}.
Extensions augment a same base language (in the case at hand, C) and
in the face of local syntax errors, assume that the current line being
parsed will be
filtered by one of the following extensions. In other words, what would
normally be regarded as severe errors is simply flushed onto
the standard output stream in our current implementation.
Obviously such
strategy is far from ideal, as it shifts all possible syntax checks
down to C compile time. A better strategy would be to let the system
try different extensions on each input parsing block, or even better
to have the system guess which extensions to apply based on the
syntactic ``signature'' of each input fragment. 
We are currently designing yet another approach, in which
``extension tags'' are used to prefix the lines that are
meant to be parsed by the corresponding extension. This implies
that in any given parsing unit only one extension will be allowed.

Our extensions are
coded in C with Lex and YACC~\cite{LeMB92} and make use of some
simple Bash shell scripts.
Some extensions were originally developed on a Windows/Cygwin
environment while more recent ones have been devised on Ubuntu Linux.
All extensions run consistently on both environments.

Each of our extensions is uniquely identified at run-time by an extension identifier---a string
in the form ``\textsf{cpm://}$e$\textsf{/}$v$'', where $e$ and $v$ are two strings
representing respectively the extension and its version number.
``Cpm'' stands for the way we currently refer to our system of extensions, 
that is ``\Cc''.

\SubSection{Assembling Components}\label{ss:combin}

\begin{figure}[t]
\includegraphics[width=0.5\textwidth]{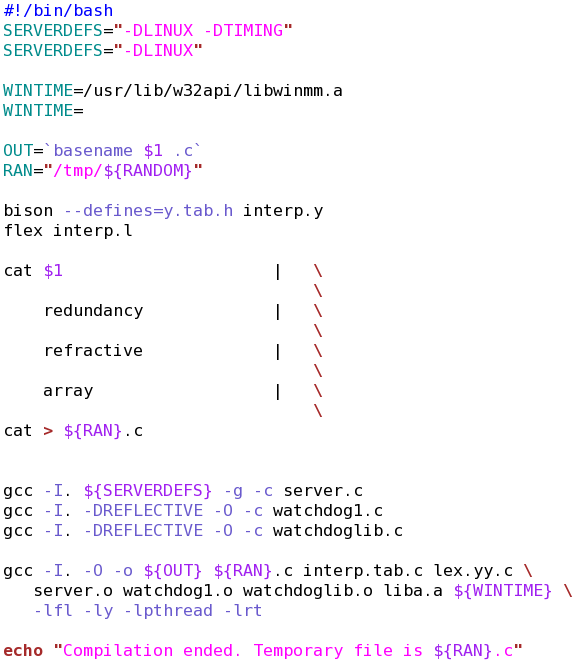}
\caption{A Bash script to apply some of the extensions and compile the resulting code.}
\label{f:compile}
\end{figure}

Our current implementation makes use of a simplistic strategy
to assemble components, requiring the user to manually
insert or remove the translators corresponding to each extension.
In particular the user is responsible for choosing the order
of application of the various extensions. Figure~\ref{f:compile}
shows the script that we use for this. A Unix pipeline is used
to represent the assembling process. Components of this pipeline are
in this case \textsf{redundancy}, which manages the extension described
in Sect.~\ref{ss:redund}, followed by \textsf{refractive}, which adds
operator overloading capabilities to context variables. The last
stage of the pipeline is in this case
\textsf{array}, which produces the extension described
in Sect.~\ref{ss:reflec}. 

It is worth pointing out that each extension publishes its extension identifier
by appending it to a context variable, a string called \textsf{extensions\_pipeline}, e.g.
``\textsf{cpm://redundancy/1.1;cpm://refractive/0.5;cpm://ar\-ray/0.5}''.
By inspecting this variable the executable is granted access to knowledge
representing the algorithmic complexity and the features of its own
execution environment.

Extensions refer to code and make use
of threads defined in libraries and ancillary programs.
Such ancillary code (and the ensuing complexity)
is then selectively loaded on demand
during the linking phase of the final compilation.

\Section{Some Preliminary Evaluation}\label{s:evalue}
In order to analyze the performance of our method we shall focus on
a particular case study: The design of a software fault-tolerant watchdog timer (WDT).
This particular choice stemmed from a number of reasons: 

\begin{itemize}
\item 
	First of all, WDT provides a well known and widespread
dependable design pattern that is often used in either hardware
or software in mission-critical embedded systems, as it provides
a cost-effective method to detect performance failures~\cite{Cri91}.

\item
	Secondly, a WDT is a real-time software. This means that it
requires
context awareness of time. This makes it suitable for being developed
with the extension described in Sect.~\ref{ss:reflec}. 

\item
	Moreover, a WDT is a cyclic application. Linguistic constructs such as the one described
in Sect.~\ref{ss:schedu} allow a concise and lean implementation of cyclic
behaviors.

\item
	Furthermore, WDT is a mission-critical tool: A faulty or hacked WDT
may cause a healthy watched component to be stopped; this in turn may severely
impact on availability. Protecting a WDT's control variables could help
tolerating some faults and security leaks. The extension
described in Sect.~\ref{ss:redund} may provide---to some extent---such
protection.

\item
	Finally, the choice of focusing on a WDT permits us to leverage
from our past research:
In~\cite{DeDe02c} we introduced a domain-specific language that permits
to define WDTs in a few lines of code.
This allows an easy comparison of the amount of the expressiveness of the two
approaches.
\end{itemize}

A context variable called \textsf{watchdog} reflects the state of
a WDT. States are reified as integers greater than $-4$. Negative values
represent conditions, i.e. either of:
  \begin{description}
    \item{\textsf{WD\_STARTED},} meaning that a WDT task is running and waiting
          for an activation message.
    \item{\textsf{WD\_ACTIVE},} stating that WDT has been
          activated and now expects periodical heartbeats from a watched task.
    \item{\textsf{WD\_FIRED},} that is, no heartbeat was received during
          the last cycle---the WDT ``fired.''
    \item{\textsf{WD\_END},} meaning that the WDT task has ended.
  \end{description}
Positive values represent how many times the WDT reset its timer
without ``firing.''

That same variable, \textsf{watchdog}, is also an actuator, as it controls the operation
of the WDT: Writing a value into it
restarts a fired WDT.

Being so crucial to the performance of the WDT, we decided to
protect \textsf{watchdog} by making it redundant. To do so
we declared it as 
\textsf{extern redundant\_t int watchdog}. Using
the \textsf{extern} keyword was necessary in order to change the \emph{definition\/}
of \textsf{watchdog} into a \emph{declaration\/}~\cite{KeRi2},
as the context aware component defines \textsf{watchdog} already.
In other words this is a practical example of two non-orthogonal extensions.

\begin{figure}[t]
\includegraphics[width=0.5\textwidth]{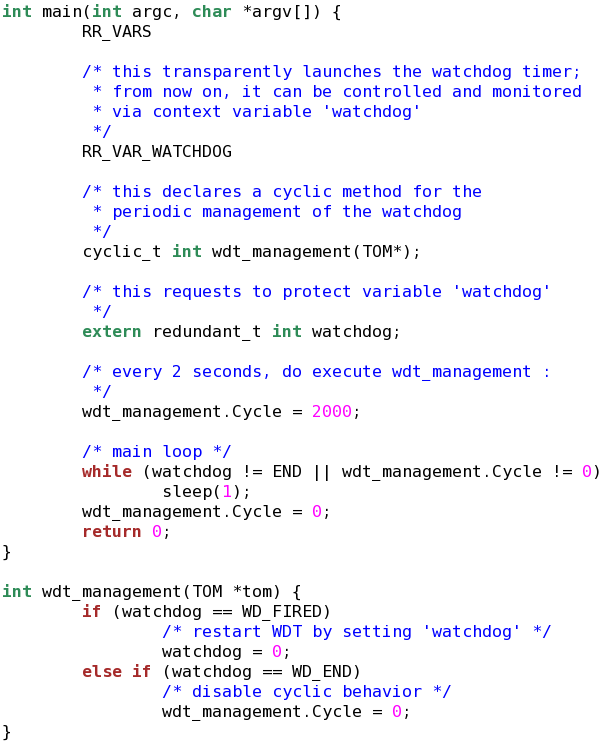}
\caption{Excerpt from the code of the WDT.}
\label{f:wdtc}
\end{figure}

Figure~\ref{f:wdtc} describes our illustrative implementation. The code uses
all three extensions reported in Sect.~\ref{s:bascom}.
A WDT thread is transparently spawned. Such thread is
monitored and controlled via variable \textsf{watchdog}. Redundant
copies of this variable are used to mitigate the effect of
transient faults or security leaks affecting memory. The code then uses our
cyclic methods extension to call periodically
a management function. Such function in turn makes use
of two of our extensions---for instance,
the WDT is restarted by writing a value into
\textsf{watchdog}.

In order to evaluate the complexity introduced by our approach we
divided complexity into an exogenous and an endogenous component,
which we call respectively
syntactical and semantic complexity.

\emph{Syntactical\/} complexity is related to the expressiveness of the language available to the
programmer. To roughly estimate this we assumed that a language's
expressiveness is inversely
proportional to the lines of code required to program with it.
If we restrict ourselves to the above discussed WDT we can
observe how in this special case the programmer is required to produce
an amount of lines of code notably lesser
than what normally expected for a comparable C program.
Such amount is slightly greater than in the case
treated in~\cite{DeDe02c}, where a C implementation of a WDT
is produced from a high level domain-specific language.
It must be remarked though that the WDT produced by our former tool
is much simpler than the one presented here---e.g. it is fault-intolerant
and context agnostic.

\emph{Semantic\/} complexity is the complexity necessary to express and make use of our extensions.
A fair and objective estimation is more difficult in this case, as it would require us to
compare different instances of our modular, loosely coupled extensions with as
many monolithic implementations. 
We have not performed such a Sisyphean task; instead, we merely observe how our approach
allows the designer to deal with a number of separated, limited problems instead of a
single, larger problem. From the divide-and-conquer design principle we then conjecture
a lesser complexity for our approach. Moreover, in our case the designer is aware 
and in full control of the amount and of the nature of the complexity he/she is adding to C. 
This in particular 
means that the programmer has fine-grained knowledge and control over the overhead 
introduced by the EE as well as over its algorithmic complexity.

\Section{Conclusions}\label{s:conclu}

We have introduced an approach to gradually augment the features of a programming
language by injecting a set of light-weighted extensions. 
Depending on the
desired features and the overhead and behaviors permitted by the target platform
and cost constraints, our approach allows the
programming language to be flexibly reshaped. 
This is because our approach employs
well-defined ``complexity containers'', each of which grants limited specific
functions and is characterized by well-defined complexity and overheads.
By doing so, complexity is made transparent
but it is not hidden: While the programmer can benefit of high level constructs, the
designer and the deployer can deal with modular building blocks each characterized by
a certain algorithmic complexity and therefore each accountable for a certain overhead.
A mechanism allows each building block to be identified, thus avoiding mismatches between
expected and provided features. At the same time, this
provides the designer with finer control
over the amount of resources required by the run-time executive of the resulting
language.

Our current implementation is merely a proof of concepts; as such it is rather
limited and in particular sacrifices elegance and efficiency to fast prototyping.
Future work will focus on improving these aspects and especially on
schemes to automatically assemble required extensions and perform more strict
analyses of syntax errors, e.g. by labeling parsing lines with ``extension tags''.
Proper GNU build system scripts will be written to automatize extension assembling
and final code compilation.
Other work will include designing a public API for third parties to
develop their own extensions. New extensions shall be designed, e.g.
one addressing parallel programming through the LINDA primitives
and inspired by our previous work~\cite{DeMS94}, and a second one
to manage ``service groups'', i.e. redundant groups of task replicas.
The latter could be used to realize constructs such as the redundant
watchdog of~\cite{DeDe02c}.

\subsection*{Acknowledgments}

We would like to thank Nicolas Letor for designing the network components of our
switchboard application and for the ideas we exchanged while designing
our context aware extension.

\bibliographystyle{latex8}
\bibliography{/lib/thesis}

\end{document}